\documentclass[aps,prl,twocolumn,showpacs]{revtex4}

\begin{document}

\title{Interferometric evidence for brane world cosmologies}

\author{Ghenadie~N.~Mardari}
\email[]{ghena@eden.rutgers.edu}

\affiliation{Rutgers University, 89 George Street, New Brunswick,
New Jersey, 08901}

\date{\today}

\begin{abstract}
The hypothesis of brane-embedded matter appears to gain increasing
credibility in astrophysics. However, it can only be truly
successful if its implications on particle interaction are
consistent with existing knowledge. This letter focuses on the
issue of optical interference, and shows that at least one
brane-world model can offer plausible interpretations for both
Young's double-slit experiment, and the experiments that fit less
neatly with it. The basic assumption is that particles can
interact at a distance through the vibrations induced by their
motion on the brane. The qualitative analysis of this mechanism
suggests that fringe visibility in single photon interference
depends on the energy levels and the interval between interacting
particles. A double-slit experiment, performed with coherent
single red photons should reveal the disappearance of interference
when the time delay between individual particles is increased over
$2.18$ seconds. In the case of infrared photons with the frequency
of $9\cdot 10^{13} Hz$, interference must vanish already at the
interval of one second.
\end{abstract}

%\keywords{}

\maketitle

Analyses of recent galaxy clustering data and CMB measurements
demand special explanations for the apparent flatness of the
Universe, as well as for its accelerated expansion. Initial
attempts to provide a simple answer in terms of vacuum energy ran
into well-known theoretical difficulties. Quantum field theories
yielded unrealistic values for the cosmological constant
\cite{wein,sahn}, while various solutions to the problem required
additional explanatory variables. In contrast, many of the most
recent brane world cosmologies are offering explanations that
appear to be both theoretically promising and empirically
testable. Most of these involve scenarios with matter localized on
flat domain walls (3-branes), embedded in higher dimensional bulk
space-times. Following the successful general formulation of
Randall and Sundrum \cite{rand,sund}, brane-world cosmologies were
applied to many problems, including the cosmological constant
\cite{stye}, cosmological phase-transitions \cite{perk}, inflation
\cite{dval,khou}, baryogenesis \cite{gaba}. Consistency with
cosmological observations has also been suggested
\cite{deff,alca}. It should be also noted, however, that the idea
of matter localized on a brane cannot be entirely satisfactory
theoretically, unless it is also shown to have plausible
implications for small scale phenomena. (How can it be essential
for large scale dynamics, and absent on smaller scales?). In other
words, it must also have quantum-level implications that would be
consistent with existing interpretations in terms of non-classic
principles, such as non-locality, uncertainty, complementarity.
However, if it were indeed possible to show that a brane model
could offer a full explanation of particle interactions, then
those mentioned principles would become redundant. The purpose of
this letter is to show that at least one such model is possible,
and that its implications can be verified with optical
interferometric equipment.

Few other tests in physics are as conceptually important as the
double-slit experiment. Because it reveals so clearly the
non-classical properties of matter, it is one of the cornerstones
of quantum mechanics. The essence of the mystery of double-slit
interferometry is that individual photons are emitted and detected
as particles, yet they seem to pass through the two slits as waves
(non-locality). Any attempt to detect the trajectory of photons
destroys the interference fringes on the detecting screen. This is
the basis for the conclusion that which-path information and
fringe visibility are complementary, just like the information
about momentum and position of particles \cite{pdds,trif}. The
evidence in favor of this interpretation is almost overwhelming,
and does not need to be listed here. Nevertheless, there are two
special cases that do not exactly fit into the mainstream picture.
For reasons that will become obvious later on, these exceptions
need to be presented here in brief.

One of these odd cases is the interference of particles emitted by
independent sources \cite{paul}. Just like in the double-slit
experiment, photons produced by independent sources produce
interference fringes, even when only one of them is present in the
system at the same time. This phenomenon was initially
demonstrated by Pfleegor and Mandel in the 1960's \cite{pfle}, and
conclusively confirmed by Hariharan \textit{et al.} in 1993
\cite{hari}. The problem with this phenomenon is that it provides
partial which-path information without destroying interference.
Initially, an interlocking mechanism between the two sources was
assumed, in the attempt to suggest that which-path information
would still be unavailable. Subsequent experiments with optically
isolated photon sources ruled out this interpretation \cite{otto}.
However, even if quantum processes washed which-path information
away, the fact remains that each photon was only emitted by one
source. This clearly shows that individual photons do not need to
go through more than one slit at the same time in order to produce
interference. Moreover, the fringes were only present when at
least two sources were active. This opens up the possibility that
photons might not interfere with themselves, as commonly assumed.
They could also affect the trajectory of subsequent particles
through some hidden mechanism.

The other odd case, related to single-photon interference, refers
to the possibility that interference might disappear entirely at
extremely low emission rates of photons. It was reported in the
1960's by Dontsov and Baz \cite{dbaz}, who obtained reliable
evidence in favor of this effect, but had no means to detect its
parameters with high precision. This experiment is in clear
contradiction with existing theories and seems to stand alone in
its field. However, the hypothesis of delayed interaction between
photons, suggested in the previous paragraph, also requires the
disappearance of interference beyond a certain interval between
individual particles. So, either the hypothesis is wrong, or there
is something special about the suggested hidden variable that
might explain with greater success these findings. Thus, we need a
model that could explain not only the classical double-slit
experiment, but also its ''anomalies``. In light of the above, the
most straightforward way to conceive of such a model is by
assuming that the brane itself is the hidden variable.

Recall that the turning point in the development of brane-world
cosmologies was the realization that a 3-dimensional domain wall
could be embedded in a bulk with extra non-compactified dimensions
[3,4]. But the same reasoning can be applied to a lower level: one
dimensional infinite strings with perfect elasticity could be
embedded in the domain wall. Accordingly, longitudinal waves
propagating on the strings would be expected to leak into the
domain wall at a rate that is inversely related to the distance of
propagation, while transverse waves would leak according to the
inverse square law. The entire brane (actually, a superstructure
of such strings) would have to be precisely tuned, so that both
s-waves and p-waves would propagate at the speed of light.
Correspondingly, particles of matter would be moving through the
brane by exciting an infinite number of strings. The collective
effect of such excitations would amount to the production of
fields that are capable to affect other particles on the same
brane. (This provides an alternative to virtual particles).

In conformity with such a model, the elastic collisions of
particles and their remote interactions are assumed to be mediated
by the oscillating brane. The photons (assumed here to be
particles) represent a special case, because they usually appear
to produce no forces. This can be explained through the following
considerations. First, photons travel in space at the speed of
light, which is also the speed of wave propagation in the brane.
Hence, their effect on the brane cannot precede them. Second, they
must produce equal amounts of symmetric waves in the plane that is
perpendicular to their direction of motion. Thus, photons can be
neutral, while still having a real effect on the brane. Third,
their effect in reverse must correspond to that of a very narrow
electrostatic beam, but it should be extremely weak, because of
the Doppler effect. Moreover, this force has to decay as it
propagates, according to the inverse square law. Consequently, the
spatial effect of photons must be temporary, and it can only be
effective in deflecting the trajectory of other photons that cross
their trails. When the trajectory of a photon is significantly
bent under the influence of an external force, its lateral
neutrality must be disrupted. Consequently, photons can produce
other forces in special environments, as detected e.g. in the case
of the photoelectric effect.

Despite the qualitative nature of this description, it can be used
to make estimates of the maximum duration of the stipulated
delayed photonic effects. Essentially, we need to calculate the
distance $(r)$ travelled by a transverse wave on a string, until
enough of its energy leaks into the brane to disable its effect on
other photons. Dividing this distance by the speed of propagation,
we could get an idea about the amount of time that it takes for
these photonic trails to fade away (r/c=t). Under the assumption
that particles always interact through brane waves, the known
energy of photons ($E_{max}=h\nu$) must be equal to their maximum
effect on the spatial strings. Moreover, the impact of photons on
strings must have a constant increment value during chosen units
of time. This basic unit of action must be equal to $h$. Hence,
the total energy of the photon comes from the cumulative number of
such interactions. This also means that a string will always be
able to deflect the trajectory of a photon if its own vibrations
have an action potential comparable to $h$. Yet, when the waves on
a string have already decayed to $E'=\nu'h$, with
$\nu'=10^{-3}Hz$, the deflection of the weakest photon cannot
exceed $0.06$ degrees (small enough to eliminate interference
build-up in most experimental settings). Thus, the value of $r$
corresponds to the distance travelled by a string wave until its
energy $E_{max}$ decays to $E'$. ($r_{0}^{2}E_{max}=r^{2}E'$,
where $r_{0}$ is the unit of distance corresponding to the system
of units used to measure $E$). The equation for the corresponding
time simplifies to $t=r_{0}\sqrt{\nu}/c\sqrt{\nu'}$.

This conclusion enables us to calculate the maximum time delay
between individual coherent photons of known frequency, at which
interference fringes might still be visible. For the red photons
(the most widely used in interferometric experiments), $t=2.18$
seconds. Other visible colors can produce interference at even
larger intervals. At the same time, infrared photons with
$\nu=9\cdot 10^{13} Hz$ must loose their fringe visibility at
$t=1$ second. Hence, it is only normal for the double slit
experiment (and for settings with independently emitted coherent
photons) to produce interference fringes, even when just one
particle passes through the system at the same time, provided the gap ($t$)
is below its critical values.

Consider the following experimental setting. A strong coherent red
laser beam is divided into two jets with a beam splitter. The jets
are focused with two mirrors in front of a screen, producing
interference fringes visible to the naked eye. A disc, connected
to a source of angular motion, is placed in front of the focal
point such as to cut off both jets. The disc has just one small
hole close to its edge, and is calibrated such as to open the
access to the screen of only one jet at a time, when it is
spinning. In other words, this is an alternator switch, whose
frequency can be adjusted by controlling its angular velocity.
When the disc has a very small velocity (less than one revolution
per minute), there can be no interference. However, at very high
velocities the interference fringes should reappear on the screen,
just like in all the other experiments with independent sources.
At this point, two video cameras are introduced into the lab. One
is used to film the switch; the other is focused on the screen.
The two recordings must later be compared frame by frame, in order
to link every interference image to its beam, and every blank
screen to intermediate states. As a result, the experimenter
acquires complete which-path information without destroying
interference. The next step is to test the correlation between
energy levels and maximum gap admissible for interference. This
can be achieved by gradually accelerating the alternator, until
the moment of qualitative shift is identified. Note that this
moment does not have to correspond to the presence of sharp
fringes on the screen. Any sign of interference works for the
proposed interpretation. The experiment can be repeated with
monochromatic lasers of different colors, in order to verify the
suggested variation of maximum gap values with changes of $\nu$.

The same findings can also be verified for non-classical states of
light, even though at higher expense. A classical Young setting is
all that is necessary, but a deterministic source of single
photons is preferable. (The development of such sources was first
announced in 2000 \cite{mich,sant,loun}, but many others have been
reported since). The idea is to perform the double-slit experiment
at very large intervals between individual coherent photons
(beyond two seconds), in order to verify the equation for $t$. It
is quite difficult to obtain good fringe visibility at such gaps,
given the drift problem. However, as specified above, high image
quality is not required in this case. Any sign of interference
beyond the specified values of $t$ would falsify the theory.
However, if the first experiment proposed above turns out to be
successful, such falsification becomes highly unlikely.

As a final note, these two experiments should not be perceived as
contradictory to existing scientific data, because they are based
on non-invasive methods of path detection. Traditionally,
which-path information is obtained with filters or detectors, both
of which affect the spatio-temporal properties of photons. In
other words, these techniques have a real (negative) impact on
coherence. It is a well-established fact that coherence is a
necessary condition for interference. Therefore, previous
experiments are not reliable as tests of locality and/or realism.

%\begin{acknowledgments}
%%%%%%%%%%%%%%%%%%%%%%%%%%
%\end{acknowledgments}

%\bibliography{brnbibs}

\end{document}